\documentstyle[epsfig]{mn}

\def\bm{\mathbf}
\begin{document}
\title{CMB Polarization Data and Galactic Foregrounds: Estimation of
  Cosmological Parameters}

\author[Prunet et al.]  {S. Prunet,$^{1, 2}$ Shiv K. Sethi,$^3$ and
  F. R. Bouchet $^1$  \\
$^{1}$ Institut d'Astrophysique de Paris, CNRS, 98bis Boulevard Arago
F--75014 Paris, France\\
$^{2}$ Canadian Institute for Theoretical Astrophysics, Mc Lennan Physical
Labs, 60 St George Street, Toronto, ON M5S 3H8, Canada\\
$^{3}$ Mehta Research Institute, Chhatnag Road, Jhusi,
Allahabad - 211019, India}

\maketitle

\begin{abstract}
  We estimate the accuracy with which various cosmological parameters
  can be determined from the CMB temperature and polarization data
  when various  galactic unpolarized and polarized
  foregrounds are included and
  marginalized using the multi-frequency Wiener filtering technique.
  We use the specifications of the future CMB missions MAP and PLANCK
  for our study. Our results are in qualitative agreement
  with earlier results obtained without foregrounds, though the errors
  in most parameters are higher because of degradation of the
  extraction of polarization signal in the presence of foregrounds.
\end{abstract}

\section{Introduction}

One of the primary goals of cosmology is to accurately determine 
various cosmological parameters associated with the background FRW
universe and the  structure formation in the universe
($\Omega$, $\Omega_\Lambda$, $\Omega_B$, $h_0$, etc.).  In recent
years compelling theoretical arguments have emerged which suggest
that the study of
CMB anisotropies is the best hope to achieve this goal (Bond 1996,
Knox 1995, Jungman {\em et al. \/} 1996, Bond {\em et al. \/} 1997,
Zaldarriaga {\em et al. \/} 1997). On the experimental
front,  two forthcoming satellite experiments MAP and 
{\sc Planck}\footnote{For details  see
  {\tt http://map.gsfc.nasa.gov} and 
{\tt http://astro.estec.esa.nl/SA-general/Projects/Planck}} 
along with a series of ground-based  and balloon-borne 
experiments on degree to sub-arcminute scales
 plan to  unravel the angular  power spectrum of the CMB to 
angular scales $\ga 1'$ (for details of interferometric ground-based
experiments
see White {\em et al. \/} 1997 and references
therein; for a recent update on balloon-borne experiments see
Lee {\em et al. \/} 1999). It has been shown that an accurate
determination of the CMB temperature fluctuations down to sub-degree 
scales could fix the values of nearly 10 cosmological parameters with
unprecedented precision (Jungman {\em et al. \/} 1996).
In addition, the future satellite missions
might detect the small,  hitherto elusive signal from CMB polarization
fluctuations (Bouchet {\em et al. \/} 1999 -- hereafter Paper I -- and
references therein).  The polarization data can be  used to 
break degeneracy between a few parameters which are determined only in
a combination using the temperature data alone (Zaldarriaga {\em et
  al. \/} 1997).

One of the major difficulties in extracting the power spectrum of
temperature and polarization fluctuations of  the CMB is the presence
of galactic and extragalactic foregrounds. The extragalactic
foregrounds (radio and infra-red point sources, clusters, etc) will
only affect small angular scales ($\la 10'$) at frequencies dominated by
the CMB signal (Toffolatti {\em et al. \/} 1997, 1999).
The galactic foregrounds, on the other hand, are
present at all angular scales and are strongest on the largest
scales. They will therefore have to be cleaned
from the future data before any definitive statements about the
primary CMB signal can be made.  A multi-frequency Wiener filtering
approach was developed to study the implications of the presence of
foregrounds for the performance of future CMB missions (Bouchet
{\em et  al. \/}  1995, Tegmark \& Efstathiou 1996).
It was shown that the primary CMB temperature 
signal is much larger than the contaminating foreground for all
the  angular scales relevant for future satellite missions. And
therefore the performance of  future all-sky satellite missions in
extracting the CMB temperature power spectra is 
unlikely to be hindered by galactic foregrounds (Bouchet
{\em et  al. \/}  1995, Tegmark \& Efstathiou 1996, Gispert \& Bouchet
1996, Bouchet \& Gispert 1999a,b).  

In a previous paper (paper I)
we extended this technique to include polarization
and temperature-polarization cross-correlation of foregrounds to
estimate their
effect on the extraction of CMB  polarization power spectra. 
Our analysis showed  that the presence of foregrounds should not
seriously deter the  detection of  $E$-mode CMB polarization and $ET$
cross-correlation  by {\sc Planck}. 
 However, while the detection of CMB polarization
will be easiest at the Doppler peaks of polarization fluctuations
 $\ell \ge  100$, where it should help reducing the errors on the 
measurement of parameters that will already be well constrained by temperature
data alone, the truly new information from polarization data
in the determination of cosmological parameters is contained in
angular scales corresponding to $\ell \le 30$ (Zaldarriaga {\em et
  al.} 1997). 

The polarization data helps break degeneracy between $C_2$, the
quadruple  moment of CMB temperature fluctuation and $\tau$, the
optical depth to the last scattering surface (Zaldarriaga
{\em et al. \/} 1997). The former gives the
overall normalization of the CMB fluctuations  and is fixed at the
epoch of inflation in  inflationary paradigm. The breaking
of this degeneracy also results in a better
determination of other inflationary parameters $T/S$, the ratio of
tensor to scalar quadruple, and the tensor index $n_T$. The optical
depth to the last scattering surface is crucial to understanding the
epoch of reionization in the universe. Even a value of $\tau$ so small
as 0.05 leave a telltale signature in the CMB polarization
fluctuations which is potentially detectable (Zaldarriaga 1997).
The main difficulty in
using polarization data to break the $C_2 \hbox{--} \tau$ degeneracy
is that one needs  to use information on large angular scales. The power
spectra at small $\ell$ is not  only badly determined because of
cosmic variance but also because of largely unknown level of polarized
foregrounds.

Prunet {\em et al. \/} (1998) attempted to model the dust
polarized emission from the galaxy ---which is the dominant foreground
for {\sc Planck} HFI--- for scales between $30'$ to a few
degrees. They showed that  though one might obtain meaningful estimates for
degree scales, there can be  large uncertainties in the large scale ($\ell
\le 50$)  polarized dust emission in the galaxy. The polarized
synchrotron is the other major galactic foreground---and it is likely
to undermine the performance of MAP and {\sc Planck} LFI. For the lack of
reliable data, we assumed the power spectra of polarized synchrotron
to mimic that of the unpolarized component in Paper I. Though there
remain large uncertainties on the polarization foregrounds,  these assumed
levels of foregrounds combined with 
the Wiener filtering methods developed in Paper I allow us to
quantify the effect of foregrounds on the extraction of CMB signal.
In this paper,  we use  the methods developed
in Paper I to ascertain  the errors   in the cosmological
parameter estimation.

In the next section, we briefly review the Fisher matrix approach
used in determining  the errors on the extraction of
cosmological parameters. We
take three underlying theoretical models for our study,
the rationale for which is
briefly described in the next section. The results are presented and
discussed in $\S$ 3. In $\S$ 4 we give our conclusions,  and discuss the
various shortcomings of our approach.

\section{Fisher Matrix and Parameter Estimation}

Future CMB missions MAP and {\sc Planck} will reach pixel
sensitivities of $\simeq 30 \, \mu K$ and $\simeq 1.5 \, \mu K$,
respectively. This should allow a very precise determination of
temperature power spectrum and
a possible 
detection of the polarization fluctuations (see paper I). 
Given the noise level and
the underlying theoretical model, the Fisher matrix approach allows
one to get an estimate of the errors in the estimation of the
parameters of the underlying model. It is
defined as an average value of the second
derivatives of the logarithm of the Likelihood function 
with respect to the cosmological
parameters,  at the true parameters value (for details see Kendall
 \& Stuart 1969):
\begin{equation}
F_{ij}=\left\langle\frac{\partial^2\mathcal{L}}{\partial \theta_i
\partial\theta_j}\right\rangle_{\Theta=\Theta_0}
\end{equation}

For CMB temperature and polarization data, the Fisher matrix can be
expressed as (Tegmark {\em et al. \/} 1997):
\begin{equation}
  F_{ij} = \frac{1}{2}Tr \left [ {\cal C}^{-1} \frac{\partial \cal C}
    {\partial \theta_i} \,
    {\cal C}^{-1}  \frac{\partial \cal C}{\partial \theta_j}  \right ]
\end{equation}
where ${\cal C} $ stands for the covariance matrix of the data and
$\theta_i$  correspond to cosmological parameters.  The details of
derivation of the covariance matrix and its derivatives
in the presence of foregrounds are  given in
Appendix A.
The error in the estimation of parameters is given by:
\begin{equation}
  \Delta \theta_i = \left [F^{-1} \right ]_{ii}^{-1/2}
\end{equation}

\subsection{Underlying models}

The estimated errors on parameters will depend on the choice of
the underlying model. We consider three models for our study.
Though these models do not exhaust all the possible models and  their
variants, our aim is to understand the errors in parameter estimation
for sCDM model  and its  popular variants, within the framework of
generic inflationary models. We are interested in 
the standard parameters of flat FRW cosmology, $h$, $\Omega_B$, $\Omega_{\nu}$,
$\Omega_\Lambda$, the reionization parameter $\tau$, and the
inflationary parameters, $C_2$, $n_s$, $T/S$, and $n_T$.
 It is of course possible to consider a more general class of
inflationary models which leads to a further proliferation of
inflationary parameters (Liddle 1998, Souradeep {\em et al. \/} 1998, Kinney
{\em et al. \/} 1998, Lesgourgues {\em et al. \/} 1999).
We also do not consider open/close  universes because, as
shown in Zaldarriaga {\em et al. } (1997), in such universes
the shift in the angular size of the horizon at the last scattering
surface leaves a very significant sign in the CMB fluctuations which
cannot be mimicked by a change in any other parameter, and therefore
$\Omega_{\rm total}$ is extremely well determined for open/close
universes. It is possible for the universe to be flat (or
nearly flat) with contribution from both matter and cosmological
constant, and one could attempt to measure both these parameters from
CMB data. However, the degeneracy between these two parameters cannot
be broken by CMB data alone and one will have to resort to other
measurements like observation of
supernova at high-z to lift this degeneracy (Tegmark
{\em et al. \/} 1998, Efstathiou \& Bond 1999).

In addition, it is possible to include parameters like $n_\nu$, the number of
massless neutrinos, and $Y_{He}$, the helium fraction. However, these
parameters can be  better determined by particle physics or local
observations (Jungman {\em et al. \/} 1996,  Bond {\em et al. \/}
1997).  Parameters like $\Omega_{\nu}$, the
contribution of massive neutrinos to the rest mass density in the
universe, can be determined to a comparable accuracy by the data
of future galaxy surveys like SDSS (Hu {\em et al. \/} 1998). 

In this paper, we take only CMB data for our study and do not
include priors from other measurements like future Galaxy surveys or high-z
Supernova results. The three models we consider
are:
\begin{itemize}
  \item[1.] sCDM model with $\tau = 0.1$. The rather large value of
    $\tau$ is taken to bring out the effects of polarization
    data.
  \item[2.] Tilted CDM model with $\tau = 0.1$, $n_S = 0.9$, $n_T =
    -0.1$, and $T/S = 0.7$. Note that $T/S = -7n_T$, which is one of
    the predictions of slow-roll inflation (Starobinsky 1985, Liddle \& Lyth 1992).
  \item[3.] Model 2 with $\Omega_{\nu} = 0.3$
    with two light, massive neutrinos.
\end{itemize}

\section{Results}

We use the results of Paper I (the values of  various terms in 
the covariance matrix as defined in Appendix A)
for the specification of the future satellite
missions. Our results
are shown in Tables~\ref{scdm},~\ref{ten},~\ref{withnu}
for the three underlying models.

It should be noted that we fix the value of
$C_2 = C_2^S + C_2^T = 796 \, (\mu K)^2$
for all the models. Only for  the sCDM model does it correspond to COBE
normalization.  In the models with tensor contribution,
the COBE normalized CMB signal  is  larger than the signal for our
normalization by a factor of $\simeq 1.5$.

To assess the reliability of our code we computed the errors in the
$6$-parameters sCDM model of Zaldarriaga {\em et al.} 1997 with their 
instrumental specifications and without foregrounds, and compared our results with both
theirs and those obtained by Eisenstein {\em et al.} 1999. Our results 
are comparable to those of Eisenstein {\em et al.} 1999 for most parameters,
with the exception of $\tau$ where the error we find is noticeably bigger.
We think that this discrepency is related to the way we compute the
derivatives of the spectra with respect to the parameters (see appendix A).

The results for sCDM model are shown in Table~\ref{scdm}.  For
comparison, results for the best channel of each experiment without
foregrounds are also shown. As is clearly seen, the performance of
Wiener filtering matches the best channel case for all the
experiments. In  Paper I
we showed that the Wiener filtering performance in extracting the
temperature 
power spectra lies between the expected performances of the best
channel and the combined sensitivity of all channel for each
experiment, at least for the specific foreground models considered. 
As the temperature data
alone gives a fair idea of the 
errors on most of the parameters our results could be
anticipated from conclusions of Paper I. However, the errors of $C_2$
and $\tau$ are mostly determined by the polarization data. In paper I
we showed that the extraction of polarization power spectra is
degraded as compared to the cases with no foreground. Our results in this
paper suggest that it should not 
be too much of a deterrent in determining
cosmological parameters. 
It is also important to note that  the present results for
the best channel case are comparable
to Wiener filtering case. This means that i) the other channels can
effectively be used to clean the best channel ii) the presence  of
foregrounds do not introduce additional  degeneracies which are
absent when the data is assumed to contain only CMB and  pixel noise. 

In Table~\ref{ten}, the expected errors are shown for a model which
includes contribution from tensor modes. One of the aims of studying
this model is to establish how well the inflationary parameters can be
determined. In comparison with the sCDM case, the errors on all the
standard parameters are  bigger. This is because the additional
parameter $T/S$ allows one to fix the normalization more freely,
thereby introducing additional degeneracies (Zaldarriaga {\em et
  al. \/} 1997). The errors on  parameters  like $C_2$, $h$ and
$\Omega_B$ are  higher than for
similar models considered by Zaldarriaga {\em et al. \/} (1997). This
is partly due to our choice of normalization which gives smaller
signal. However, it also reflects the degradation of the polarization
power spectra extraction in the presence of foregrounds. Other
parameters like $T/S$, $\tau$, and $n_T$ are better determined  than the
results of  Zaldarriaga {\em et al. \/} (1997), but it is
mostly  owing to the fact that we take larger input values for $\tau$
and $T/S$. We also show the effect of including very small signal
from $B$-mode polarization. As is seen, it results in a better
determination of most parameters, especially the inflationary
parameters. Though the $B$-mode signal is much smaller  compared to
$E$-mode signal, and is generally below the pixel noise except for
a small range of modes for $\ell \la 100$, its very presence indicates
tensor modes in inflationary paradigm. Also,
the degradation of the extraction of this signal in the
presence of foreground is
smaller than for the $E$-mode signal (Paper I).
Therefore, it can make a 
difference in the estimation of
parameters. Our results show that the consistency condition of slow-roll
inflation, $T/S = -7n_T$,  can be checked by future missions (it should be 
noted here that this relation was imposed only in the fiducial model, but
excursions of both parameters were considered independently). {\sc Planck}
can extract both these parameters  with $1\sigma$ errors $\la 50 \%$.
However, it should be kept in mind that our results are more
optimistic than the results of Zaldarriaga {\em et al. \/} (1997)
because of our choice of input model.

The results  of adding another parameter $\Omega_\nu$ in the model above
are  shown in Table~\ref{withnu}. $\Omega_\nu$ can be determined to an
accuracy $\la 10 \%$ with {\sc Planck}, though it will be very difficult for
MAP to determine it. Note that errors on other
parameters have not changed much by the addition of this parameter,
which suggests that no new degeneracies have cropped up. However,
degeneracies between various parameters depend very sensitively on the
choice of input model. For instance, if the new parameter $\Omega_\nu$
was added with the input value $\Omega_\nu = 0$, it would have  substantially
worsened the estimation of almost all parameters, especially the
inflationary parameters.
For all the three models considered here, we took $\Omega_\Lambda =
0$. A finite value of $\Omega_\Lambda$  results in a better
estimation of all the parameters of FRW model as well as
$\Omega_\Lambda$ (Zaldarriaga {\em et al. \/} 1997). 

\section{Conclusions and Summary}

In this paper we 
estimated 
the effect of foregrounds on the
determination of cosmological parameters.
The most important result 
is that although the presence of foregrounds somewhat worsens the 
parameter estimation by degrading the detection of polarization
signal, it does not give rise to
any  severe degeneracies not already present in the CMB data (CMB
signal and pixel noise)
without foregrounds. It needs to be
further confirmed with a detailed study of the Likelihood function
in the multiple  parameter space. 

Any  analysis such as ours can only give a qualitative idea on the
accuracy of parameter estimation. This is largely because of its strong
dependence on the input model (Zaldarriaga {\em et al.\/} 1997). In
addition, there are great uncertainties in the assumed level of
foregrounds which we take in the Wiener filtering analysis of paper I.
Moreover, the foreground characteristics (power spectra, frequency dependence)
should be determined from the data as well, and this adds uncertainty to
the determination of the cosmological parameters (see Knox 1999).
It should be noticed at this point that, if Wiener filtering assumes some
frequency dependences as well as power spectra for the CMB and the foregrounds,
it also gives a measure of the error on the estimation of the power spectra
of foregrounds from the filtered data, see Paper I. \footnote{ It sould also be noticed
that any spatial change of the frequency indexes (for dust or synchrotron)
should correspond to special {\em astrophysical} regions (molecular clouds,
supernovae remnants\ldots) and that an analysis where any such spatial change
of index is simply incorporated as an additional ``noise'' term would lead to 
pessimistic results. This would rather point out that a global analysis
(in ``Fourier'' modes) is insufficient to take this effect properly into
account.} 

Still our results  suggest that the primary obstacles for high precision CMB
measurements will rather stem from systematic errors and inaccuracies in
calibration, baseline drifts, determinations of far side lobes or
estimates of filter transmissions\ldots all of which are of course not
included in this analysis. Furthermore, a Fisher matrix analysis leads
to the smallest possible error bars, which can only be degraded by
inaccuracies in devising the Wiener filters (by using approximate power
spectra and spectral behaviours). 

The next step will be to directly analyse 
simulated mega-pixel multi-frequency CMB maps relevant to future
experiments. However, such an analysis is an extremely difficult
(if not intractable) numerical
problem (for recent attempts see  Muciaccia {\em et al. \/} 1997, Oh
{\em et al. \/} 1999, Borrill 1999). 
In light of this, our results should be regarded  as a first
attempt on the problem of parameter estimation in the presence of
foregrounds, which give a qualitative idea  of the expected accuracy
in parameter estimation
till the analyses of multi-frequency CMB datasets become possible.
Since the submission of this work, more detailed analysis of the effect
of foregrounds have been investigated by Tegmark {\em et al.\/} 1999. 
Their results are similar to ours, with maybe slightly higher foreground
residuals as they allowed some scatter in the frequency dependence of 
foregrounds.   

\section{Appendix A}

In this section we briefly recapitulate the discussion of paper I, and
derive the covariance matrix of CMB data and its derivatives.
The observed CMB data at multiple frequencies can be expressed in
multipole space as:
 
\begin{equation}
y_{\nu}^i (l,m)=A_{\nu p}^{ij} (l,m) x_p^j (l,m) + b_{\nu}^i (l,m)
\label{filmod}
\end{equation}
where $x_p^j$ is the underlying signal for process $p$ and ``field'' $j$ (i.e.
temperature or (E,B) polarization modes), and $\nu$ is a frequency channel index.
In the Wiener filtering method, one considers a linear relation
between the true, underlying signal, $x_p^j$ and the linearly optimal
estimator of 
the  signal,  $\hat x_p^j$. 
\begin{equation}
  \hat x_p^i = W_{p\nu}^{ij} y_{\nu}^j.
  \label{fildef}
\end{equation}
Eqs.~(\ref{filmod}) and (\ref{fildef}) can be used to write the
estimated power spectrum as:
\begin{eqnarray}
  \langle \hat x_p^i  \hat x_{p'}^j \rangle &=& ({\bm WA})_{p p''}^{im}
  ({\bm WA})_{p' p'''}^{jq} \langle x_{p''}^m   x_{p'''}^q \rangle\nonumber\\
  &&+ W_{p\nu}^{il} W_{p'\nu'}^{jn} \langle b_\nu^l b_{\nu'}^n \rangle \nonumber\\
  &\equiv& Q_{pp'}^{ij} \langle  x_p^i   x_{p'}^j \rangle
  \label{filpow}
\end{eqnarray}
where the last equality comes from the equation defining the Wiener filter
(see Eq.~$6$ of Paper I). 
The covariance of the  filtered data  can then be written as:
\begin{equation}
 {\cal C_{\ell}}  = \left(\begin{tabular}{ccc}
      $Q^{11}_{\ell} C_{T\ell}$ & $ Q^{12}_{\ell} C_{TE\ell}$ & $0$ \\
      $Q^{12}_{\ell} C_{TE\ell}$  & $Q^{22}_{\ell} C_{E\ell}$ & $0$ \\
      $0$ & $0$ & $Q^{33}_{\ell} C_{B\ell}$ \\
    \end{tabular}\right )
\end{equation}

For computing the Fisher matrix we also need to compute the derivative
of  the covariance with respect to cosmological parameters:
\begin{equation}
  \frac{\partial {\cal C_{\ell}}}{\partial \theta_i} = \sum_{X = T, E,
    ET,B}\frac{\partial {\cal  C_{\ell}}} {\partial C_{\ell}(X)}
\frac{\partial {C_{\ell}(X)} }{\partial \theta_i}
\end{equation}
The derivative of the covariance matrix with respect to various
power spectra can be written using Eq.~(\ref{filpow}). These 
derivatives, it should be borne in mind, are with respect to the
 {\em input power spectra \/ } used in estimating the Fisher matrix
and not the power spectra used in constructing the Wiener
filters, which, therefore,  are invariant  under this change. These
derivatives can be readily calculated:
\begin{eqnarray}
  \frac{\partial \langle \hat x_p^T  x_p^T  \rangle}{\partial C_p^T} 
  &=& 
  (W_{p\nu}^{11} A_{\nu p}^{11})^2 \\ 
  \frac{\partial \langle \hat x_p^T  x_p^T  \rangle}{\partial C_p^{TE}} 
  & = & 
  2 \times (W_{p\nu}^{11} A_{\nu p}^{11}W_{p\nu}^{12} A_{\nu p}^{22}) \\
  \frac{\partial \langle \hat x_p^T  x_p^T  \rangle}{\partial C_p^E} 
  &=& 
  (W_{p\nu}^{12} A_{\nu p}^{22})^2 \\
  \frac{\partial \langle \hat x_p^E  x_p^E  \rangle}{\partial C_p^E} &
  = & 
  (W_{p\nu}^{22} A_{\nu p}^{22})^2 \\
  \frac{\partial \langle \hat x_p^E  x_p^E  \rangle}{\partial C_p^{TE}} 
  & = & 
  2 \times (W_{p\nu}^{22} A_{\nu p}^{22}W_{p\nu}^{21} A_{\nu p}^{11})
  \\
  \frac{\partial \langle \hat x_p^E  x_p^E  \rangle}{\partial C_p^T} &
  = & 
  (W_{p\nu}^{21} A_{\nu p}^{11})^2 \\
  \frac{\partial \langle \hat x_p^T  x_p^E  \rangle}{\partial C_p^{T}} 
  & = & 
   (W_{p\nu}^{11} A_{\nu p}^{11}W_{p\nu}^{21} A_{\nu p}^{11}) \\
  \frac{\partial \langle \hat x_p^T  x_p^E  \rangle}{\partial C_p^{TE}} 
  & = & 
   (W_{p\nu}^{11} A_{\nu p}^{11}W_{p\nu}^{22} A_{\nu p}^{22})  +
   (W_{p\nu}^{12} A_{\nu p}^{22}W_{p\nu}^{21} A_{\nu p}^{11}) \\
   \frac{\partial \langle \hat x_p^T  x_p^E  \rangle}{\partial C_p^{E}} 
  & = & 
   (W_{p\nu}^{22} A_{\nu p}^{22}W_{p\nu}^{12} A_{\nu p}^{22}) \\
 \end{eqnarray}

 Theoretical power spectra are calculated using the CMB Boltzmann code
 CMBFAST (Seljak  \& Zaldarriaga 1996). Derivatives with respect to
 cosmological parameters are calculated numerically using a variant of
 {\tt dfridr} routine of numerical recipes (Press {\em et al. \/}
 1992).
 We notice that a $5 \%$
 step in most parameters gives stable results. The only exception is
 derivative of $E$-mode power spectra
 with respect to $\tau$ when $\tau \le 0.05$ for $\ell \le 20$. This
 numerical instability is expected as a small change  in this
 parameter when the  input value of $\tau$ is very small
 can cause  appreciable
 change in the $E$-mode power spectra at small $\ell$. However, the
 numerical differentiation is quite stable for $\tau \ga 0.05$.

\clearpage

\begin{table}
  \caption{Errors on parameters for sCDM model with $\tau = 0.1$. Also
    shown are the corresponding errors for the best channel of each experiment.}
  \label{scdm}
{\begin{tabular}{|l|c|c|c|c|c|c|}
\hline
\hline
Parameters & $C_2$ & $h$ & $\Omega_b$ & $\Omega_{\Lambda}$ & $\tau$ &  $n_S$\\
\hline
Model & $796 (\mu K)^2$ & $0.5$ & $0.05$ & $0.0$ & $0.1$ & $1.0$\\
\hline
Wiener ({\sc Planck}) & {$2.4$ \%} & {$1.36$ \%} & {$2.3$ \%} & $0.039$ &
{$4.6$ \%} & {$0.34$ \%}\\
\hline
Best Channel ({\sc Planck})  & {$2.1$ \%} & {$1.06$ \%} & {$1.82$ \%} & $0.03$ & {$3.74$\%} & {$0.3$ \%}\\
\hline
Wiener  (HFI) & {$2.48$ \%} & {$1.39$ \%} & {$2.37$ \%} & $0.04$ & {$5.75$
  \%} & {$0.35$ \%}\\
\hline
Best channel (HFI) & {$2.1$ \%} & {$1.06$ \%} & {$1.83$ \%} & $0.03$ &
{$3.75$ \%} & {$0.3$ \%}\\
\hline
Wiener (LFI) & {$3.81$ \%} & {$2.26$ \%} & {$3.72$ \%} & $0.067$ & {$10.3$
  \%} & {$0.54$ \%}\\
\hline
Best Channel(LFI) & {$3.6$ \%} & {$2$ \%} & {$3.3$ \%} & $0.057$ & {$6.5$
  \%} & {$0.51$ \%}\\
\hline
Wiener (MAP) & {$4.9$ \%} & {$4$ \%} & {$8.9$ \%} & $0.12$ & {$45.6$
  \%} & {$1.65$ \%}\\
\hline
Best Channel(MAP) & {$6.3$ \%} & {$4.5$ \%} & {$10.7$ \%} & $0.13$ & {$43.5$
  \%} & {$1.76$ \%}\\
\hline
\hline
\end{tabular}}
\end{table}

\medskip

\begin{table}
  \caption{Errors on parameters for a model with tensor
    contribution with or without the inclusion of $B$-mode
    polarization}
  \label{ten}
{\begin{tabular}{|l|c|c|c|c|c|c|c|c|}
\hline
\hline
Parameters & $C_2$ & $h$ & $\Omega_b$ & $\Omega_{\Lambda}$ & $\tau$ &
$n_S$ & $n_T$ & $T/S$\\
\hline
Model & $796 (\mu K)^2$ & $0.5$ & $0.05$ & $0.0$ & $0.1$ & $0.9$ &
$-0.1$ & $0.7$ \\
\hline
Wiener ({\sc Planck}) & {$8.7$ \%} & {$1.6$ \%} & {$2.7$ \%} & $0.045$
& {$5.5$\%} & {$0.46$ \%} &{$81$ \%} & {$22.4$ \%} \\
\hline
{+} B-modes ({\sc Planck})  & {$6.5$ \%} & {$1.55$ \%} & {$2.64$ \%} & $0.044$
& {$4.8$\%} & {$0.43$ \%} &{$57.1$ \%} & {$17.5$ \%} \\
\hline
Wiener  (HFI) & {$9.4$ \%} & {$1.63$ \%} & {$2.8$ \%} & $0.05$ & {$7$
  \%} & {$0.47$ \%} &{$87$ \%} &  {$24.1$ \%} \\
\hline
{+} B-modes (HFI) & {$7.7$ \%} & {$1.6$ \%} & {$2.7$ \%} & $0.05$ & {$6$
  \%} & {$0.45$ \%} &{$70$ \%} &  {$20.6$ \%} \\
\hline
Wiener (LFI) & {$9.8$ \%} & {$5.3$ \%} & {$8.6$ \%} & $0.15$ & {$11.3$
  \%} & {$1.65$ \%} &{$91.6$ \%} & {$32.4$ \%}  \\
\hline
{+} B-modes(LFI) & {$9$ \%} & {$4.6$ \%} & {$7.5$ \%} & $0.13$ & {$9.6$
  \%} & {$1.42$ \%} &{$83$ \%} &  {$28.2$ \%} \\
\hline
Wiener (MAP) & {$12.3$ \%} & {$22.3$ \%} & {$40$ \%} & $0.67$ & {$52.5$
  \%} & {$7.5$ \%} & {$91$ \%} & {$91$ \%}\\
\hline
{+} B-modes (MAP) & {$12$ \%} & {$20$ \%} & {$36.5$ \%} & $0.60$ & {$46$
  \%} & {$7$ \%} & {$90.4$ \%} & {$81.6$ \%}\\
\hline
\hline
\end{tabular}}
\end{table}

\medskip

\begin{table}
  \caption{Parameter estimation with $\Omega_\nu$}
  \label{withnu}
{\begin{tabular}{|l|c|c|c|c|c|c|c|c|c|}
\hline
\hline
Parameters & $C_2$ & $h$ & $\Omega_b$ & $\Omega_{\Lambda}$ & $\tau$ &
$n_S$ & $n_T$ & $T/S$ & $\Omega_{\nu}$ \\
\hline
Model & $796 (\mu K)^2$ & $0.5$ & $0.05$ & $0.0$ & $0.1$ & $0.9$ &
$-0.1$ & $0.7$ & $0.3$\\
\hline
Wiener ({\sc Planck}) & {$8.8$ \%} & {$1.6$ \%} & {$3$ \%} & $0.045$
& {$4.6$\%} & {$0.95$ \%} &{$82$ \%} & {$24$ \%} & {$10.5$ \%} \\
\hline
{+} B-modes ({\sc Planck})  & {$6.4$ \%} & {$1.45$ \%} & {$2.65$ \%} & $0.04$
& {$4.3$\%} & {$0.81$ \%} &{$55$ \%} & {$18$ \%} &{$9.45$ \%}   \\
\hline
Wiener  (HFI) & {$9.6$ \%} & {$1.66$ \%} & {$3.1$ \%} & $0.045$ & {$6$
  \%} & {$0.97$ \%} &{$89$ \%} &  {$26$ \%} & {$10.8$ \%}\\
\hline
{+} B-modes (HFI) & {$7.75$ \%} & {$1.55$ \%} & {$2.9$ \%} & $0.42$ & {$5.5$
  \%} & {$0.9$ \%} &{$71$ \%} &  {$22.6$ \%} & {$10$ \%} \\
\hline
Wiener (LFI) & {$9.2$ \%} & {$3.6$ \%} & {$6$ \%} & $0.098$ & {$11$
  \%} & {$2$ \%} &{$84$ \%} & {$31$ \%} & {$26.5$ \%} \\
\hline
{+} B-modes(LFI) & {$9.2$ \%} & {$3.6$ \%} & {$6$ \%} & $0.097$ & {$10.8$
  \%} & {$2$ \%} &{$84.8$ \%} &  {$31$ \%} & {$26.5$ \%}  \\
\hline
Wiener (MAP) & {$12.3$ \%} & {$18.4$ \%} & {$27$ \%} & $0.69$ & {$67.1$
  \%} & {$7.5$ \%} & {$115$ \%} & {$75$ \%} & {$201$ \%}\\
\hline
{+} B-modes (MAP) & {$12$ \%} & {$17$ \%} & {$25.4$ \%} & $0.65$ & {$62$
  \%} & {$7.1$ \%} & {$115$ \%} & {$70$ \%} & {$200$ \%}\\
\hline
\hline
\end{tabular}}
\end{table}

\begin{thebibliography}{}  
\bibitem[]{} Bennett C. L., {\em et al. \/} 1996, ApJ, 464, L1
\bibitem[]{} Bond J. R., Efstathiou G., \& Tegmark M. 1997, MNRAS, 291, 33
\bibitem[]{} Bond J. R. 1996, Observations of Large-Scale Structure in the
universe, Ed. Schaeffer R., Les Houches, Elsevier Science Publishers
\bibitem[]{} {Bouchet} F.~R., {Gispert} R., \& {Puget} J.-L. 1995, In
``Unveiling the Cosmic Infrared Background'', AIP Conference Proceedings 348,
Baltimore, Maryland, USA, E.~{Dwek}, editor, pages 255--268
\bibitem[]{} Borrill J. 1999, Phys. Rev. D, 59, 027302
\bibitem[]{} Bouchet F. R., Prunet S. \& Sethi S. K. 1999, MNRAS, 302, 663
\bibitem[]{} Bouchet F. R. \& Gispert R. 1999a, submitted to {\it New
Astronomy}, astroph/9903176
\bibitem[]{} Bouchet F. R. \& Gispert R. 1999b, in preparation
\bibitem[]{} Bouchet F. R., Gispert R., Aghanim N., Bond J. R.,
De Luca A., Hivon E., Maffei B. 1995, Space Sci. Rev., 74, 37
\bibitem[]{} Efstathiou G.  \& Bond J. R. 1999, MNRAS, 304, 75
\bibitem[]{} Eisenstein D. J., Hu W. \& Tegmark M., 1999, ApJ, 518, 2
\bibitem[]{} Gispert R. \& Bouchet F. R. 1996, Proceedings of the $16^{th}$
Rencontres de Moriond, ``Microwave Background Anisotropies'', 
eds.  F. Bouchet, R. Gispert, B. Guiderdoni, \& J. Tr\^an Thanh
V\^an, Editions Fronti\`eres, p?? 
\bibitem[]{} Hu W.,  Eisenstein D. A. \& Tegmark M. 1998, Phys. Rev. Lett., 80,
5255
\bibitem[]{} Jungman G., Kamionkowski M., Kosowsky A., Spergel D. N. 1996,
  Phys. Rev. D, 54, 1332
\bibitem[]{} Kamionkowski M. \& Kosowsky A. 1998, Phys. Rev. D, 57, 685
\bibitem[]{} Keating B., Timbie P., Polnarev A., \& Steinberger, J. 1998,
 ApJ, 495, 580
\bibitem[]{} Kendall M. G. \& Stuart A. 1969, {\em The Advanced}
   {\em Theory of Statistics}, Vol. II, Griffin, London
\bibitem[]{} Kinney W. H., Dodelson S. \& Kolb, E. W. 1998, in Proc. of the 
33th Rencontres de Moriond, ``Fundamental Parameters in Cosmology'', eds 
J. Tr\^an Thanh V\^an, Y. Giraud-H\'eraud, F. Bouchet, T. Damour \& Y. Mellier,
Editions Fronti\`eres, p3
\bibitem[]{} Knox L. 1995, Phys. Rev. D, 52, 4307
\bibitem[]{} Knox L. 1999, MNRAS, 307, 977
\bibitem[]{} Lee, A. T. {\em et al. \/} 1999, astro-ph/9903249
\bibitem[]{} Lesgourgues J., Prunet S. \& Polarski D. 1999, MNRAS, 303, 45
\bibitem[]{} Liddle A. R. \& Lyth D. H. 1992, Phys. Lett. B, 291, 391
\bibitem[]{} Liddle A. 1998, in Proc. of the 
33th Rencontres de Moriond, ``Fundamental Parameters in Cosmology'', eds 
J. Tr\^an Thanh V\^an, Y. Giraud-H\'eraud, F. Bouchet, T. Damour \& Y. Mellier,
Editions Fronti\`eres, p81
\bibitem[]{} Lineweaver C. H. \& Barbosa D., 1998, ApJ, 496, 624L; see also
  http://www.sns.ias.edu/~max/cmb/experiments.html
\bibitem[]{} Muciaccia P. F., Natoli P., \& Vittorio N. 1998, ApJ,
  488, L63
\bibitem[]{} Oh S. P., Spergel D. N., \& Hinshaw G. 1999, ApJ, 510, 551
\bibitem[]{} Prunet S., Sethi S. K., Bouchet F. R.,
  \& M. -A. Miville--Desch\^enes, 1998, A\&A, 339, 187
\bibitem[]{} Seljak U. \& Zaldarriaga M. 1996, ApJ, 488, 1
\bibitem[]{} Smoot G., {\em et al. \/} 1992, ApJ, 396, L1
\bibitem[]{} Souradeep T., Bond J. R. 1998, to appear in the
  proceedings of COSMO-97, Ambleside, England
\bibitem[]{} Starobinsky A. A., 1985, Pis'ma Astron, Zh., 11, 323
\bibitem[]{} Tegmark M., Eisenstein D. A. \& Hu W. 1998, Proc. of the 33th
Rencontres de Moriond, {\it Fundamental Parameters in Cosmology}, eds J. Tr\^an Thanh
V\^an, Y. Giraud-H\'eraud, F. Bouchet, T. Damour, Y. Mellier, Editions Fronti\`eres,
p355
\bibitem[]{} Tegmark M., Taylor A. \& Heavens A. 1997 ApJ, 480, 22
\bibitem[]{} Tegmark M. \& Efstathiou G. 1996, MNRAS, 281, 1297
\bibitem[]{} Tegmark M., Eisenstein D. J., Hu W., de Oliveira-Costa A., 
astro-ph/9905257
\bibitem[]{} Toffolatti L., Argueso Gomez F., De Zotti G., 
Mazzei P., Franchescini A., Danese L., Burigana C. 1998, MNRAS, 297, 117
\bibitem[]{} White M., Carlstorm J. E., \& Dragovan M. 1997,
  Apj in press 
\bibitem[]{} Zaldarriaga M., Spergel D. N., Seljak U. 1997, ApJ, 488, 1
\bibitem[]{} Zaldarriaga M. 1997, Phys. Rev. D, 55, 822
\end{thebibliography}
\end{document}